%% file: main.tex
\documentclass[
]{ceurart}

\usepackage{listings}
\usepackage{array}
\usepackage{booktabs}
\usepackage[normalem]{ulem}
\usepackage{multirow}  
\usepackage{xcolor}
\usepackage[shortlabels]{enumitem}

\DeclareUnicodeCharacter{2212}{-}

\begin{document}

\copyrightyear{2025}
\copyrightclause{Copyright for this paper by its authors.
  Use permitted under Creative Commons License Attribution 4.0
  International (CC BY 4.0).}

\conference{ISWC'25: International Semantic Web Conference,
  November 2--6, 2025, Nara, Japan}

\title{Accelerating Scientific Discovery with Multi-Document Summarization of Impact-Ranked Papers}

\author[1,2]{Paris Koloveas}[%
    orcid=0000−0003−2376−089X,
    email=pkoloveas@athenarc.gr
]
\address[1]{IMSI, Athena RC, Athens, Greece}
\address[2]{University of the Peloponnese, Tripolis, Greece}

\author[1]{Serafeim Chatzopoulos}[%
    orcid=0000−0003−1714−5225,
    email=schatz@athenarc.gr
]


\author[1]{Dionysis Diamantis}[%
    orcid=0009-0002-3272-1294,
    email=dionysis.diamantis@athenarc.gr
]

\author[2]{Christos Tryfonopoulos}[%
    orcid=0000−0003−0640−9088,
    email=trifon@uop.gr
]

\author[1]{Thanasis Vergoulis}[%
    orcid=0000−0003−0555−4128,
    email=vergoulis@athenarc.gr
]


\input{sections/abstract}


\maketitle

\input{sections/1_introduction}
\input{sections/2_related_work}

\input{sections/3_system_overview}
\input{sections/4_main_functionalities}
\input{sections/5_user_interface}

\input{sections/6_demonstration_scenarios}
\input{sections/7_conclusions}

\section{Acknowledgments}
This work has received funding from the EU's Horizon Europe framework programme as part of the SciLake project (GA: 101058573).

\bibliography{thebib}

\end{document}

%% file: sections/abstract.tex
\begin{abstract}
The growing volume of scientific literature makes it challenging for scientists to move from a list of papers to a synthesized understanding of a topic. Because of the constant influx of new papers on a daily basis, even if a scientist identifies a promising set of papers, they still face the tedious task of individually reading through dozens of titles and abstracts to make sense of occasionally conflicting findings. To address this critical bottleneck in the research workflow, we introduce a summarization feature to BIP! Finder, a scholarly search engine that ranks literature based on distinct impact aspects like popularity and influence. Our approach enables users to generate two types of summaries from top-ranked search results: a concise summary for an instantaneous at-a-glance comprehension and a more comprehensive literature review-style summary for greater, better-organized comprehension. This ability dynamically leverages BIP! Finder's already existing impact-based ranking and filtering features to generate context-sensitive, synthesized narratives that can significantly accelerate literature discovery and comprehension.

\end{abstract}

%% file: sections/1_introduction.tex
\section{Introduction}
\label{sec:intro}

The growth rate of scientific publications continues to accelerate, driven by an expanding global research community and the intense pressure to ``publish or perish''.
This trend has been amplified by the emergence of AI-assisted writing tools that facilitate generating publishable content \cite{Mou2024TheSF,Alzaabi2023ChatGPTAI}.
Consequently, identifying the most valuable and relevant articles for a given research query has become a challenging task. 
Academic search engines like Google Scholar\footnote{\url{https://scholar.google.com/}} and Semantic Scholar\footnote{\url{https://www.semanticscholar.org/}} attempt to mitigate this issue by combining keyword relevance with citation-based impact measures to rank results, helping researchers prioritize their reading.
However, these systems often rely on simplistic impact measures, like raw citation counts, which are prone to manipulation, biased against recent articles, and treat impact as a single dimension, overlooking the different aspects of scientific impact \cite{KanellosVSDV21}.

BIP! Finder~\cite{bip-finder} addresses this issue by incorporating multiple impact indicators that capture distinctly different aspects of a publication's impact, such as its overall impact (also known as `influence') and its current attention (also known as `popularity'). 
As a scientific search engine, BIP! Finder enables users to rank results based on the most relevant impact indicator for their specific use case, offering a customizable and refined approach to scientific discovery.
However, researchers still struggle to synthesize the core findings related to a particular question of interest from a list of prioritized articles; they should manually read abstracts to connect ideas, identify trends, and grasp the overall narrative. 
While BIP!~Finder aids users prioritize what to read, this step remains demanding and time-consuming.



This paper introduces a significant AI-assisted extension to BIP!~Finder that addresses this gap. 
We present an on-the-fly summarization functionality powered by Large Language Models (LLMs), through a Retrieval-Augmented Generation (RAG)-inspired document-grounded approach,
that enables users to generate structured, cited summaries of top-ranked articles directly within the search interface.
Our main contributions can be summarized as follows:

\begin{itemize}[nosep, topsep=1pt]
    \item \textbf{Integrated Summarization:} We integrate our summarization functionality into the scientific search workflow, leveraging BIP!~Finder's existing ranking and filtering capabilities.
    \item \textbf{Dual-Mode Synthesis:} The system automatically selects between two summary types: a concise overview (for rapid understanding) and a more extensive literature review-style summary (for in-depth analysis).
    \item \textbf{Context-Aware Generation:} By building upon BIP!~Finder's eyword-based results, ranked by different impact indicators appropriate to the desired use case, the generated summaries are not of random related articles but of a coherently prioritized set, providing a more meaningful synthesis.
\end{itemize}

We should note that the aforementioned summarization functionality is readily available to the registered users of BIP! Finder\footnote{\url{https://bip.imsi.athenarc.gr/}} within its main search user interface.\footnote{For review purposes, an anonymous test account is available: Username: test\_user, Password: aRdIckiNOrte.}



%% file: sections/2_related_work.tex
\section{Related Work}\label{sec:rel_work}

Automating the synthesis of scientific literature is a rapidly advancing field, largely driven by LLMs~\cite{genaidiscovery, luo2025llm4sr}. While early research focused on extractive methods that select salient sentences from source documents~\cite{cohansumm}, our system is fully abstractive, generating new, coherent prose, that synthesizes information, rather than merely extracting it. Similarly, recent work has focused on generating structured summaries by filling predefined templates with key information like ``methodology'' or ``key takeaways''~\cite{luistructuredsum, liu2025cs, shamsabadi2024keywords}. 
Our approach complements this by producing unstructured, narrative summaries, which offers a more flexible and readable synthesis that is not constrained by a fixed schema.

Our system is best understood within the RAG paradigm, where an LLM is grounded with a trusted corpus. CORE-GPT~\cite{coregpt}, which served as a key inspiration for our work, exemplifies this as a question-answering system that first retrieves relevant papers and then generates a cited answer based on the retrieved content. LitLLMs~\cite{litllms} explores a comprehensive pipeline for literature review generation, including retrieval and plan-based text generation. Our work implements a similar philosophy but with a crucial distinction in its integration: unlike systems that programmatically generate search queries or retrieve papers for a given abstract, our summarization is performed on a set of articles that has already been intelligently filtered and ranked \textit{by the user} based on multi-faceted impact criteria.
Finally, in contrast to manually curated knowledge exploration tools like TL;DR PROGRESS~\cite{syed2024tldr}, our system is fully automated and general-purpose, capable of summarizing articles on any topic in real-time.

%% file: sections/3_system_overview.tex
\section{System Overview}\label{sec:system_overview}

\begin{figure}[t]
    \centering
    \includegraphics[width=.95\textwidth]{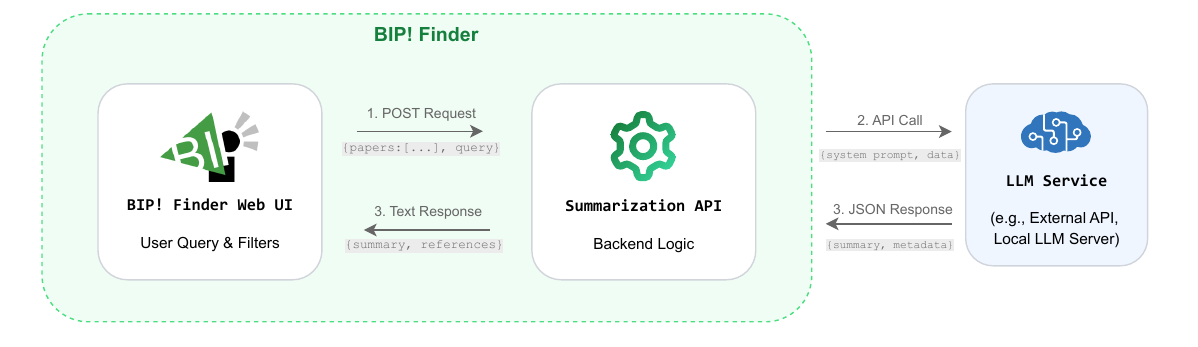}
    \caption{System Architecture of the BIP! Finder Summarization Functionality.}
    \label{fig:system_architecture}
\end{figure}

The new summarization functionality is architecturally decoupled from the main BIP!~Finder front-end to ensure modularity and scalability. The overall system architecture, depicted in Figure \ref{fig:system_architecture}, illustrates the interaction between the user-facing application, our new backend service, and the LLM service.

The process starts when the user submits 
a summarization request from the BIP!~Finder Web UI. 
The front-end sends a POST request to our Summarization API, with a payload containing the user's query as well as the list of selected articles (including their IDs, titles, and abstracts). 
Based on this, the Summarization API automatically determines the appropriate system prompt based on the number of articles and constructs a request for the LLM service; this can be either an external LLM accessed through an API or a locally deployed model.
Our backend then calls the LLM (currently DeepSeek V3~\cite{deepseekv3} is used) sending the appropriate data and the selected prompt. 
Then, the LLM processes the information and returns the generated, structured summary to our Summarization API, which finally relays it back to the BIP!~Finder Web UI to be presented to the user.



This three-tier architecture allows us to separate concerns: the BIP! Finder front-end handles user interaction and data presentation, the Summarization API manages logic and communication with the LLM service, which provides the core natural language generation capability.

The source code for the Summarization API is publicly available on GitHub\footnote{\url{https://github.com/athenarc/bip-scientific-summarization}}. The API is architected as a standalone microservice, allowing it to be deployed and used independently of the BIP!~Finder front-end; any system can utilize this service by following the request examples provided\footnote{\url{https://github.com/athenarc/scientific-summarization-api/blob/main/curl_example.md}}. 
Furthermore, the service is designed for flexibility in its choice of LLM. While our current implementation leverages the DeepSeek V3 model, the backend is compatible with any LLM that adheres to the OpenAI API standard. This allows for easy substitution with other commercial models or even locally hosted open-source models served through inference servers like TGI\footnote{\url{https://github.com/huggingface/text-generation-inference}} or vLLM~\cite{vllm}.

%% file: sections/4_main_functionalities.tex
\section{Summarization Functionality}\label{sec:main_functionalities}

The primary goal of our new feature is to accelerate knowledge synthesis. It achieves this by providing on-demand, structured summaries tailored to the user's immediate needs. The functionality is governed by two carefully engineered system prompts\footnote{\url{https://github.com/athenarc/scientific-summarization-api/blob/main/system_prompts.yaml}} that guide an LLM to produce high-quality academic text.

The system offers two modes of summarization, automatically selected based on the number of articles the user chooses to include in the summary:

\begin{itemize}[nosep, topsep=1pt]
    \item \textbf{Concise Summary (1-5 Articles):} When a user requests a summary for a small set of top-ranked articles, the system generates a tightly focused, single-paragraph summary. The goal is to provide a quick, digestible overview of the key contributions and themes. This is powered by our \texttt{conscise} system prompt.
    
    \item \textbf{Literature Review-Style Summary (6-20 Articles):} For users wanting a deeper understanding of a topic, the system can generate a multi-paragraph summary resembling a literature review. This mode is invoked when more than 6 articles are selected. 
    Our \texttt{lit-review} system prompt directs the model to structure the output into 3-4 paragraphs, starting with a contextual introduction, thematically grouping findings, and concluding with a synthesis of 
    major trends.
\end{itemize}

The quality and structure of the generated summaries are ensured through meticulous prompt engineering. Rather than sending a generic instruction, our prompts enforce a set of rules designed to produce text that is useful in an academic context.
Key instructions common to both prompts include:

\begin{itemize}[nosep, topsep=1pt]
    \item \textbf{Verifiability through Citation:} The most critical instruction is the mandatory citation for every claim using a numeric format. This transforms the summary from a simple block of text into a navigable index of the source literature, allowing researchers to immediately trace a statement back to the paper it originated from.
    \item \textbf{Grounded Generation:} By strictly forbidding the use of any information not explicitly present in the provided titles and abstracts, we mitigate the risk of LLM ``hallucinations'' or fabrication. This ensures the summary is a faithful synthesis of the provided source material.
    \item \textbf{Enforced Narrative Structure:} Both prompts dictate a specific narrative flow. For instance, the \texttt{lit-review} prompt explicitly requires an introductory paragraph that sets the context, body paragraphs that thematically group studies, and a concluding paragraph that synthesizes findings. This moves the output beyond a simple list of facts to a coherent, analytical text that highlights relationships, trends, and knowledge gaps between the papers.
    \item \textbf{Scholarly Tone and Style:} The prompts guide the LLM to adopt formal academic language, encouraging it to compare and contrast methodologies and identify consensus or contradictions, which are key elements of a genuine literature review.
\end{itemize}

%% file: sections/5_user_interface.tex
\section{User Interface}\label{sec:user_interface}

\begin{figure}[t]
    \centering
    \includegraphics[width=.92\textwidth]{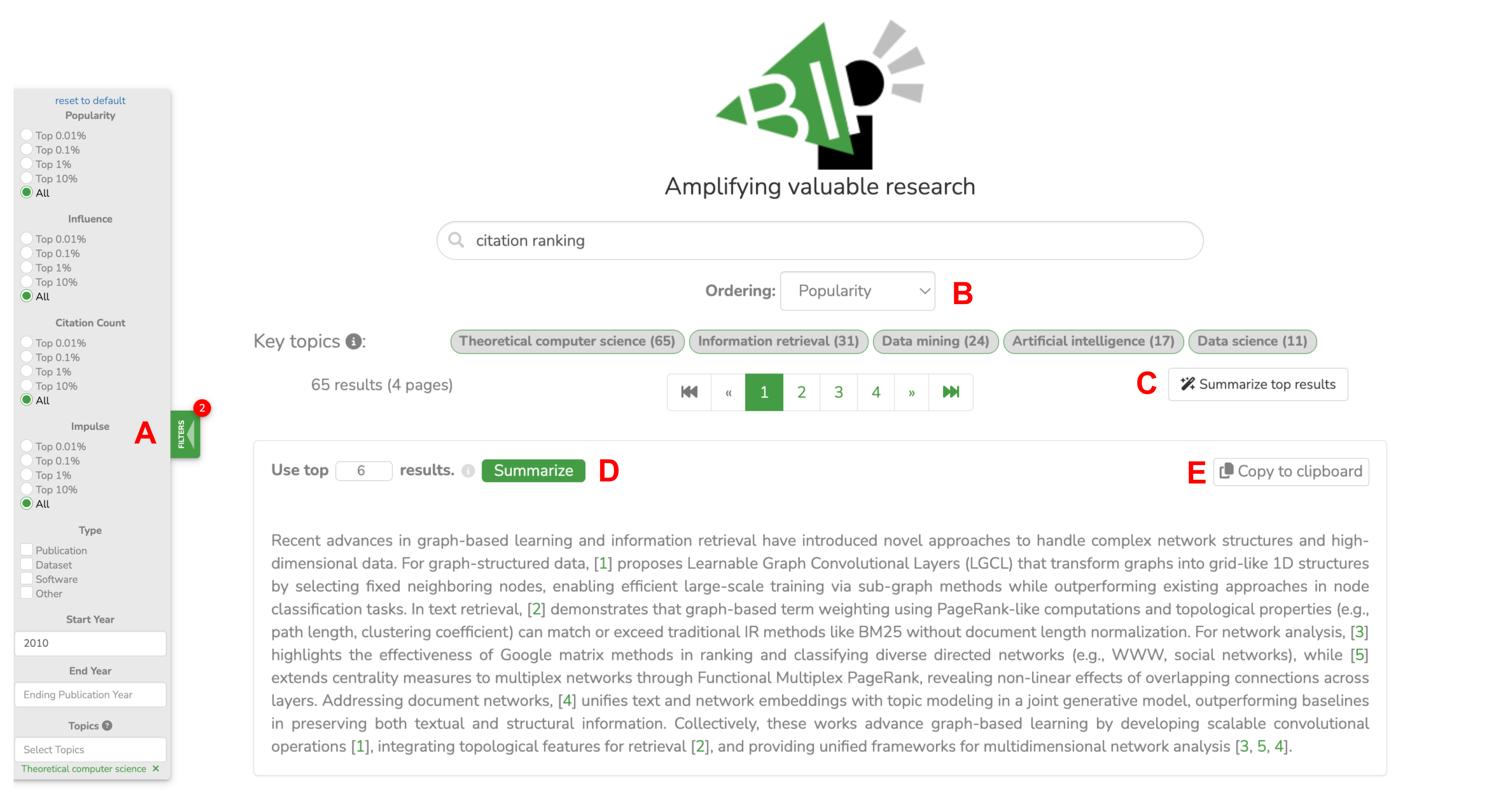}
    \caption{BIP! Finder Web UI with Summarization. 
    \textbf{A.} Filters for narrowing results. \textbf{B.} Dropdown for ordering articles by impact indicators. \textbf{C.} Button to generate a summary of top 5 articles. \textbf{D.} Options to regenerate summary with user-defined article count. \textbf{E.} Button to copy summary and references to system clipboard.}
    \label{fig:UI}
\end{figure}

The summarization functionality is integrated into BIP! Finder's search interface to support a natural user workflow, where a researcher first identifies a set of source articles before synthesizing their content. The entire process and its corresponding interface elements are visualized in Figure~\ref{fig:UI}.

A user’s interaction is designed to first shape the list of articles from which the summary will be generated. The quality of any summary is directly dependent on the relevance of its source material, so the interface provides a suite of tools for this essential curation process. For instance, a user can apply filters from the sidebar to narrow the results by a specific publication date range, focus only on certain artifact types like \textit{Publications} or \textit{Datasets}, or isolate papers belonging to one or more research topics (Figure~\ref{fig:UI}, A). In parallel, they can use the ``Ordering'' dropdown menu (Figure~\ref{fig:UI}, B) to rank these filtered results not just chronologically, but by distinct aspects of scientific impact, such as long-term \textit{Influence} or current \textit{Popularity}. 
This combination of granular filtering and sophisticated impact-based ranking allows a researcher to precisely define a high-quality, thematically coherent set of articles before proceeding to the summarization step.

Once the user is satisfied with the curated list, they can generate a synthesis. The initial summarization can be triggered by the \textbf{``Summarize top results''} button (Figure~\ref{fig:UI}, C), which uses the top 5 articles by default. For more fine-grained control, a set of controls is presented directly above the summary output area. Here, the user can adjust the number of articles to include (from 1 to 20) and click the \textbf{``Summarize''} button to produce a new summary based on the specified number (Figure~\ref{fig:UI}, D). This single adjustment directly controls the summarization mode: selecting 1-5 articles will always produce the concise summary, while selecting 6 or more articles automatically triggers the generation of the more extensive, literature review-style summary.  To facilitate the use of this generated text in the researcher's own writing or note-taking tools, a \textbf{``Copy to clipboard''} button is also provided within the summary box (Figure~\ref{fig:UI}, E), allowing for one-click transfer of the summary.

%% file: sections/6_demonstration_scenarios.tex
\section{Demonstration Scenarios}\label{sec:demonstration_scenarios}
To illustrate the capabilities of the new summarization functionality, we present two demonstration scenarios that showcase its practical applications in academic research.

\subsection{Scenario 1: Getting a Quick Gist of a Trending Topic}

An experienced researcher wants to quickly understand the latest developments in ``artificial intelligence in agriculture''.

\begin{enumerate}[nosep, topsep=1pt]
    \item The user enters the query \textit{``artificial intelligence agriculture''} into the search box.
    \item They leave the ordering on the default, ``Popularity'', to see the most recent, trending works.
    \item Instead of reading through the abstracts of the top papers one by one, they click the ``Summarize top results'' button.
    \item The system generates a concise, single-paragraph summary of the top 5 most popular articles, displayed in a text box. The summary synthesizes the key approaches and findings, with citations linking back to the articles in the results list. This allows them to grasp the current state of research in under a minute.
\end{enumerate}

\subsection{Scenario 2: Building a Foundation for a Literature Review}
A new PhD student needs to start a literature review on ``citation ranking'' and wants to identify foundational and key contributing papers.

\begin{enumerate}[nosep, topsep=1pt]
    \item The user searches for \textit{``citation ranking''}.
    \item To find the most foundational works, they change the ranking criterion from ``Popularity'' to ``Influence''. This re-orders the results to prioritize seminal papers with long-term impact.
    \item They also use the filters on the left to narrow the results to the topic ``Artificial intelligence'' to exclude irrelevant papers.
    \item They want a more detailed overview, so they use the controls above the summary to set the number of results to 10 and click ``Summarize''.
    \item The system generates a multi-paragraph, literature review-style summary. The summary introduces the core problem, thematically groups different approaches to citation ranking found in the papers, compares their methodologies, and provides a synthesized conclusion, with all claims properly cited. This provides them with a solid, structured starting point for their detailed review.
\end{enumerate}

%% file: sections/7_conclusions.tex
\section{Conclusions}

We have presented the summarization functionality implemented in the BIP! Finder scientific search engine. By using on-the-fly, dual-mode summarization in combination with state-of-the-art, impact-based ranking, our system enables researchers to accelerate the knowledge discovery process. It complements the gap between the determination of pertinent literature and synthesizing its main concepts. With this addition, we show that context-aware summarization of influence or popularity-ranked articles is well-positioned to be a solution for both overview and in-depth literature research.

